\documentclass[useAMS,usenatbib]{mn2e}

\usepackage{graphicx}
\long\def\symbolfootnote[#1]#2{\begingroup%
\def\thefootnote{\fnsymbol{footnote}}\footnote[#1]{#2}\endgroup}

\title[Stellar population gradients]{Stellar population gradients in early-type cluster galaxies with VIMOS IFU\thanks{Based on observations collected at the European Organisation for Astronomical Research in the Southern Hemisphere, Chile (run: 078.B-0539)}}
\author[T. D. Rawle et al.]{T. D. Rawle\thanks{E-mail:
t.d.rawle@durham.ac.uk}, Russell J. Smith, J. R. Lucey, A. M. Swinbank \\
Department of Physics, Durham University, Durham DH1 3LE, United Kingdom}

\begin{document}

\date{2008 July 21}

\pagerange{\pageref{firstpage}--\pageref{lastpage}} \pubyear{2008}

\maketitle

\label{firstpage}

\begin{abstract}
We present results from a pilot study of radial stellar population trends in early-type galaxies using the VLT VIMOS integral field unit (IFU). We observe twelve galaxies in the cluster Abell 3389 (z $\approx$ 0.027). For each galaxy, we measure 22 line-strength indices in multiple radial bins out to at least the effective radius. We derive stellar population parameters using a grid inversion technique, and calculate the radial gradients in age, metallcity and $\alpha$-abundance. Generally, the galaxies in our sample have flat radial trends in age and [$\alpha$/Fe], but negative gradients in [Z/H] (--0.20 $\pm$ 0.05 dex). Combining our targets with two similar, long-slit studies to increase sample size, we find that the gradients are not correlated with the central velocity dispersion or $K$-band luminosity (both proxies for galaxy mass). However, we find that the age and metallicity gradients are both anti-correlated with their respective central values (to $>$ 4$\,\sigma$), such that galaxies with young cores have steeper positive age gradients, and those with metal-rich centres have strong negative [Z/H] gradients. 
\end{abstract}


\section{Introduction}
\label{sec:intro}

Stellar population gradients offer a valuable probe of the star formation and chemical evolution of early-type galaxies. Simple models of classic monolithic collapse \citep{lar74-585,car84-403} can be ruled out by the relative weakness of the observed metallicity gradients \citep[e.g.][]{tam04-617}. Generally, age gradients are expected to be small for the short collapse times required for $\alpha$-enhancement \citep[e.g.][]{tho99-655}, while the $\alpha$-enhancement itself could show positive or negative gradients \citep*{pip07-0706.2932}. Within the hierarchical $\Lambda$CDM models \citep[e.g.][]{col00-168,del06-499,bow06-645}, merger gas content and feedback prescriptions can be investigated \citep{kaw03-908,kob04-740}; mergers would tend to dilute or erase Z/H gradients \citep[e.g.][]{whi80-1}, although dissipation could trigger star formation at the core, creating age gradients \citep{bar91-65,mih94-47}.

Traditionally, observational studies of radial trends in early-type galaxies have employed long-slit spectroscopy in local cluster and field environments (e.g. \citealt*{gor90-217,car93-553}; \citealt{meh03-423}; \citealt*{san06-823,nor06-815}; \citealt{red07-1772,san07-759}). In general, most studies find flat age gradients, small $\alpha$-abundance gradients (either positive or negative), and large negative radial trends in metallicity.

Long-slit spectrographs generally have a good optical throughput. However, the spatial data is immediately reduced to one-dimension and is dependent on the slit orientation, with most of the light from the galaxy undetected. The use of an integral field unit (IFU), which enables both spectral and spatial coverage, allows the analysis of the entire galaxy. With careful treatment, light from the entire circumference can be binned to give radial trends unbiased by a slit orientation.

One notable use of an IFU in this field is the Spectrographic Areal Unit for Research on Optical Nebulae (SAURON) project \citep{bac01-23,dez02-513}. The instrument has good spatial resolution allowing the creation of maps for each measured parameter, particularly in kinematics. However, the spectral range of SAURON is limited to four useful line indices (H$\beta$, Fe5015, Mgb5177, Fe5270$_{\rm s}$). From spatial maps of the iron lines, \citet{kun06-497} deduce that the metallicity gradients are generally negative and are roughly aligned with the light profiles. The H$\beta$ maps are typically flat, with a strong central peak in a few galaxies, interpreted as recent star formation activity in the core. Stellar population maps for the SAURON early-type spiral sample are analysed in \citet{pel07-445}.

\begin{figure*}
\includegraphics[viewport=0mm 0mm 278mm 93mm,width=165mm,angle=0,clip]{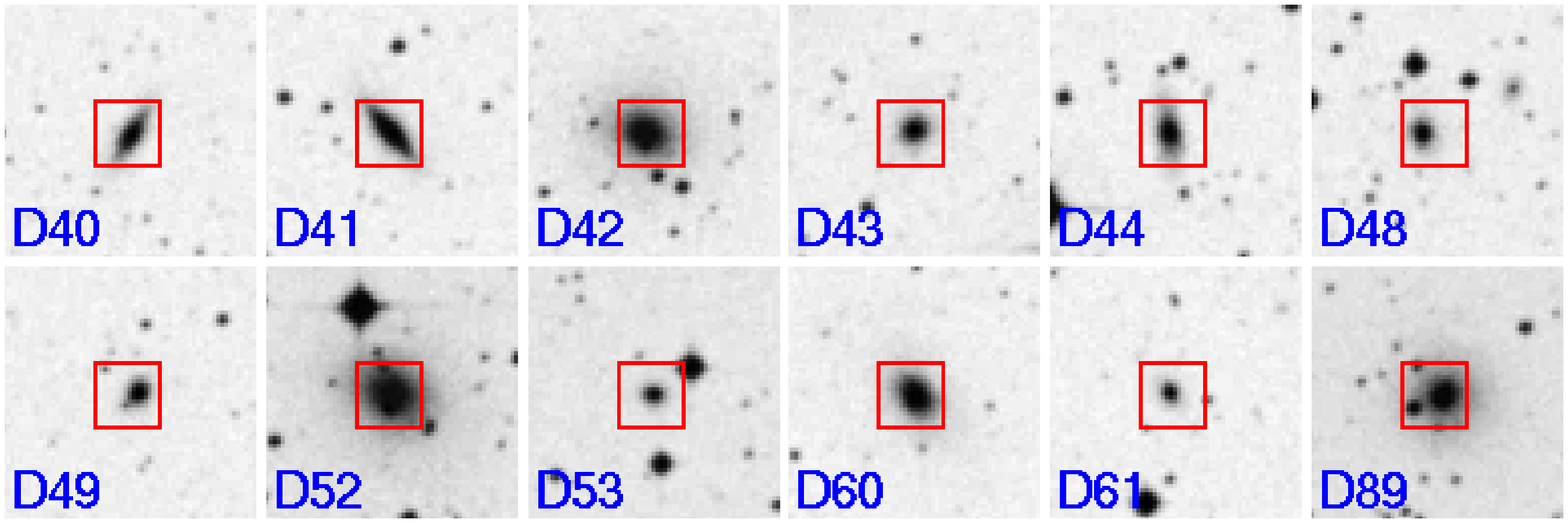}
\caption{Thumbnail images from the Digitized Sky Survey, showing a 100 arcsec square field around the target galaxies. The boxes correspond to the 27$\times$27 arcsec VIMOS field of view, underlining how well matched our chosen galaxies are to the instrument.}
\label{fig:targets}
\end{figure*}

\begin{table*}
\centering
\caption{Parameters of the target galaxies in Abell 3389. ID and morphology from \citet[][D80]{dre80-565}. Positions from 2MASS. $R$-band magnitude from NED; $K$-band magnitude from 2MASS. Effective circular radius ($r_{\rm e}$) calculated from 2MASS $J$-band images. Central velocity dispersion, $\sigma$ (km s$^{-1}$), as measured in Section \ref{sec:sigma} using the $r_{\rm e}$/3 aperture.} 
\label{tab:targets} 
\begin{tabular}{@{}clccrrrlr} 
\hline
ID & \multicolumn{1}{c}{NGC} & RA & Dec & \multicolumn{1}{c}{m$_R$} & \multicolumn{1}{c}{m$_K$} & \multicolumn{1}{c}{$r_{\rm e}$} & \multicolumn{1}{c}{Morph.} & \multicolumn{1}{c}{$\sigma$} \\
(D80) & \multicolumn{1}{c}{\#} & (J2000) & (J2000) & \multicolumn{1}{c}{mag} & \multicolumn{1}{c}{mag} & arcsec & \multicolumn{1}{c}{(D80)} & \multicolumn{1}{c}{(km s$^{-1}$)} \\
\hline
D40 & & 06 21 59.7 & --64 59 19 & 14.2 & 11.16 & 4.2 & S0 & 178.8 $\pm$ 6.1 \\
D41 & 2233 & 06 21 40.1 & --65 02 00 & 13.5 & 10.47 & 4.6 & S0 & 184.6 $\pm$ 2.4 \\
D42 & 2230 & 06 21 27.5 & --64 59 34 & 12.6 & 9.66 & 10.5 & E & 253.9 $\pm$ 3.1 \\
D43 & & 06 21 13.9 & --65 00 58 & 14.6 & 11.58 & 3.1 & S0 & 149.8 $\pm$ 4.4 \\
D44 & & 06 20 54.4 & --65 01 08 & 14.0 & 11.13 & 3.8 & S0 & 221.5 $\pm$ 4.5 \\
D48 & & 06 23 49.0 & --64 57 15 & 14.8 & 11.94 & 2.9 & E & 144.6 $\pm$ 2.3 \\
D49 & & 06 23 07.4 & --64 55 50 & 14.7 & 11.64 & 2.3 & E & 180.9 $\pm$ 3.9 \\
D52 & 2235 & 06 22 22.1 & --64 56 03 & 12.4 & 9.35 & 12.3 & D & 252.1 $\pm$ 2.4 \\
D53 & & 06 22 04.8 & --64 57 36 & 15.1 & 12.18 & 2.8 & E & 113.5 $\pm$ 4.4 \\
D60 & & 06 22 19.8 & --64 14 04 & 13.5 & 10.58 & 4.5 & E & 194.0 $\pm$ 1.9 \\
D61 & & 06 21 58.3 & --64 53 44 & 15.3 & 12.37 & 2.1 & S0 & 146.6 $\pm$ 6.7 \\
D89 & 2228 & 06 21 15.6 & --64 27 32 & 12.8 & 9.79 & 9.3 & SB0 & 247.7 $\pm$ 2.8 \\
\hline 
\end{tabular}
\end{table*} 

Here, we assess the feasibility of using the VLT VIsible MultiObject Spectrograph (VIMOS) IFU for studying radial trends of stellar populations by observing a small sample of early-type galaxies in the nearby cluster Abell 3389 (z $\approx$ 0.027). With a large future study in mind, we develop the observing strategy and analysis techniques, comparing our results with those of previous work. The paper is organised as follows: Section \ref{sec:data} describes the sample, observations and data reduction; Section \ref{sec:analysis} details the measurement of the gradients in the absorption lines and in the stellar populations; Section \ref{sec:sp_grads} analyses the stellar population gradients; Section \ref{sec:conclusion} highlights our main conclusions.

\section{Observations and data reduction}
\label{sec:data}

\subsection{The sample and observations}
\label{sec:sample}

We observed a sample of 12 early-type galaxies from the nearby cluster Abell 3389 (cz $\approx 8100$ km s$^{-1}$). This study was designed as a feasibility assessment for the methodology and as such, the target cluster was selected based on its low-demand RA ($\sim 6^h$). Each galaxy is a confirmed cluster member and is morphologically classified as an early-type by \citet[][D80]{dre80-565}. A thumbnail image from the Digitized Sky Survey\footnote{http://archive.stsci.edu/dss/} is displayed in Figure \ref{fig:targets} for each target, while Table \ref{tab:targets} lists various fundamental parameters.

The sample spans a range of three magnitudes in luminosity, and the galaxies have effective radii of 2.1 -- 12.5 arcsec (1.1 -- 6.7 kpc). The effective radii ($r_{\rm e}$) of the galaxies were measured from 2MASS $J$-band images. Due to the relatively small size of the target galaxies with respect to the 2MASS PSF (FWHM $\sim$ 3 arcsec), the apparent half-light radii required a correction. For this, we adopted the difference between the half-light radii of Sersic model fits (using {\sc galfit}; \citealt{pen02-266}) before and after PSF convolution. Our smallest object (D61) has an apparent half-light radius of 2.9 arcsec, which was corrected to 2.1 arcsec. The uncertainty in this correction, derived from the scatter in corrections for similar sized galaxies, is $\pm$0.1 arcsec. For larger galaxies, the PSF correction is smaller, and we estimate that the listed effective radii have an uncertainty of better than 10 per cent.

Ten of the twelve target galaxies were previously observed through 2 arcsec diameter fibres as part of the NOAO Fundamental Plane Survey (NFPS; \citealt{smi04-1558}, \citealt{nel05-137}). From the NFPS spectra, the only target to display line emission is D52 ($EW$[O {\sc iii} $\lambda$5007] = 0.55 \AA). D52 is a bright galaxy with a `D'-type morphology, so the emission is most likely LINER-like rather than due to star formation. The emission line ratio [O{\sc iii}]/H$\beta$ is greater than 3 for LINER-like objects and H$\gamma$/H$\beta$ $\sim$ 0.6, as inferred from nebular emission spectra. Therefore, in our worst-case galaxy, emission decreases the H$\beta$ and H$\gamma$F absorption indices by no more than 0.2 and 0.1 \AA{} respectively. Using the \citet{tho03-897,tho04-19} stellar population model grids, we find that this causes an over-prediction in the ages of $<$ 5 per cent ($<$ 1 Gyr for D52).

The target galaxies were observed in October and November 2006 with the VIsible MultiObject Spectrograph (VIMOS; \citealt{lef03-1670}) on the ESO VLT Melipal at Paranal. VIMOS was used in the Integral Field Unit (IFU) mode with the high resolution, blue grism. Table \ref{tab:vimos} details the instrument configuration used. Observations were made during dark time, with an average seeing of 0.7 arcsec (FWHM). Each galaxy was observed for 1730 seconds, achieving an average signal-to-noise ratio (SNR) $\sim$ 18 \AA$^{-1}$ at 5000 \AA, for a 1 arcsec annulus at the effective radius ($r_{\rm e}$). An offset sky frame with an exposure of 120 s was obtained for each target. Including overheads, a 1 hour observing block was devoted to each galaxy.

\begin{table}
\centering
\caption{VIMOS instrument configuration.} 
\label{tab:vimos} 
\begin{tabular}{@{}lr} 
\hline
Mode & IFU\\
Grism & HR blue\\
Field of view & 27 $\times$ 27 arcsec\\
Spatial resolution & 0.67 arcsec/fibre\\
Wavelength range & 4150 -- 6200 \AA\\
Spectral resolution & 2.1 \AA{} FWHM \\
Dispersion & 0.51 \AA/pixel\\
\hline 
\end{tabular}
\end{table} 

\subsection{Data reduction and radial binning}
\label{sec:reduction}

\begin{figure*}
\includegraphics[viewport=99mm 25mm 172mm 240mm,height=165mm,angle=90,clip]{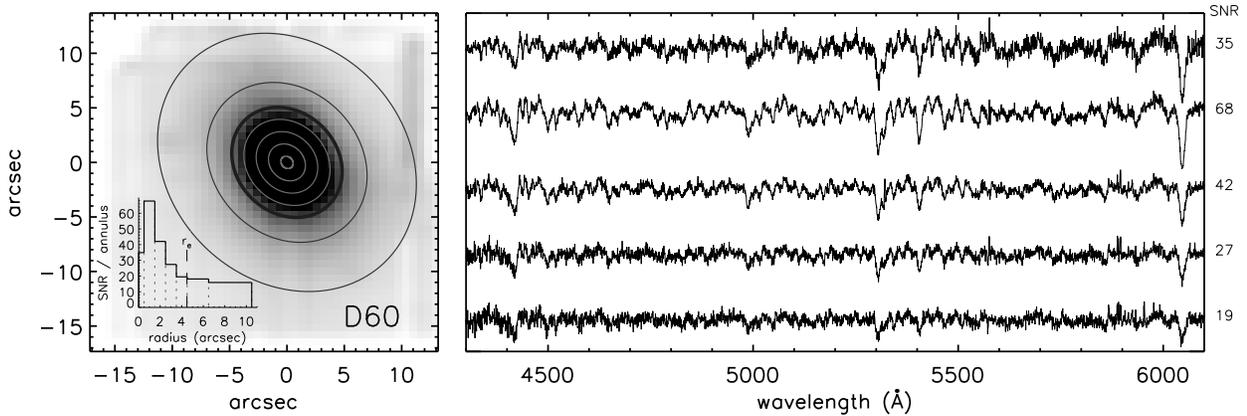}
\caption{Left panel: example of elliptical bins, showing broadband, wavelength-collapsed image from VIMOS for D60 with the adopted boundaries. Thick ellipse indicates the effective radius ($r_{\rm e}$). Inset: total SNR within each annulus; $r_{\rm e}$ shown as a dashed line. Right: spectra in the observed frame, with arbitrary offsets, for the five annuli within $r_{\rm e}$; innermost annulus at the top.}
\label{fig:bins}
\end{figure*}

Initial data reduction, from raw detector output to sky-subtracted data-cube, was achieved using the ESO common pipeline ({\sc EsoRex}) and the {\sc iraf} routine {\sc imcombine}. Each observation consists of three dithered exposures split into four quadrants, with an additional single exposure of four quadrants for the sky offset frame. For each quadrant, a sky background spectrum was calculated by taking the median over all spatial pixels in the offset exposure and then normalising, pixel-by-pixel, to the 5577 \AA{} line of the science frame. These sky spectra were then subtracted from the appropriate quadrants before performing a median combine of the three exposures. The resulting data-cubes were flux calibrated using the observed standard stars.

For each galaxy, an ellipse was fit to the broadband, wavelength-collapsed (`white') image. This ellipse formed the basis for the pseudo-isophotic annuli\footnote{Named `pseudo-isophotic' to stress that the annuli do not follow the true isophotes, but are based on the ellipse fit to the global galaxy shape.}. Working from the centre of the galaxy outwards (in steps of 0.5 arcsec), individual fibre spectra were summed in elliptical annuli, with the bin boundaries based on two criteria:

\begin{enumerate}
\item annulus SNR $\ge$ 18 \AA$^{-1}$
\item annulus width (along major axis) $\ge$ 1 arcsec
\end{enumerate}

Two exceptions were adopted: the central bin is allowed a 0.5 arcsec semi-major axis provided that the minimum SNR is achieved; the outermost bin may have an SNR $<$ 18 \AA$^{-1}$ if the radius is limited by the field of view. The targets have between three and seven radial bins, depending on effective radius and surface brightness. As an illustration, Figure \ref{fig:bins} (left panel) shows the adopted bins for D60, alongside a histogram of the resulting SNRs (inset). For the same galaxy, total spectra from the five annuli within $r_{\rm e}$ are given in the right panel of Figure \ref{fig:bins}. Additionally, we computed the spectra within three circular apertures:

\begin{enumerate}
\item 1 arcsec radius (for direct comparison to NFPS)
\item effective radius, $r_{\rm e}$
\item $r_{\rm e}/3$ (for galaxy-to-galaxy comparison)
\end{enumerate}

\subsection{Velocity dispersion}
\label{sec:sigma}

The stellar kinematics ($V$, $\sigma$) of each galaxy were computed from the spectra of each bin, using {\sc ppxf} \citep{cap04-138} and the \citet{vaz99-224} evolutionary stellar population templates. These templates cover a similar wavelength range to VIMOS (we use $\sim$4900--5500 \AA{} for fitting) and importantly, have a higher spectral sampling (FWHM $=$ 1.8 \AA) than VIMOS HR blue. Errors were calculated using 51 Monte Carlo simulations, with the 1$\sigma$ intervals taken from the 8th and 42nd values after numerically sorting.

Comparing velocity dispersion measurements from the 1 arcsec radius circular aperture to NFPS, we find a good agreement with a median offset of 0.023 $\pm$ 0.008 dex (new values generally larger), and a scatter around the offset of 0.019 dex. The scale also shows reasonable consistency, with a slope of 1.02 $\pm$ 0.06.

\begin{figure}
\includegraphics[viewport=23mm 49mm 95mm 219mm,height=84mm,angle=90,clip]{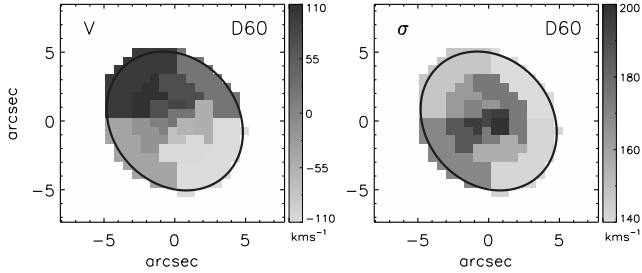}
\caption{Radial velocity (V; left panel) and velocity dispersion ($\sigma$; right panel) maps for the centre of D60 (effective radius indicated by the black ellipse). Both kinematic quantities in km s$^{-1}$, with the scale shown to the right of each panel.}
\label{fig:kin_maps}
\end{figure}

For the galaxy with the highest total SNR within $r_{\rm e}$ (D60), we created maps of the kinematics ($V$ and $\sigma$). Optimising for both SNR and spatial resolution, the data were re-binned into 16 segments of annuli, ranging from SNR$\sim$20--35 in the centre, to SNR$\sim$10--15 at the effective radius. The maps for $V$ and $\sigma$ (Figure \ref{fig:kin_maps}) show that D60 has major axis rotation of $\sim$110 kms$^{-1}$. Unfortunately, the total SNRs within $r_{\rm e}$ for our other targets were insufficient for a similar analysis. A one hour exposure with VIMOS is generally unsuited to the spatial mapping of galaxies at our target redshift (previous studies of 2D absorption line kinematics, e.g. SAURON, observed galaxies an order of magnitude closer).  The remainder of this paper adopts the radial binning described in Section \ref{sec:reduction}.

\subsection{Line strength measurements}
\label{sec:indexf}

Absorption-line strengths were measured from the flux-calibrated spectra at the VIMOS instrument resolution ($\sim$2.1 \AA{} FWHM; $\sim$ 50 km s$^{-1}$) using {\sc indexf}\footnote{http://www.ucm.es/info/Astrof/software/indexf/}. Uncertainties were calculated by {\sc indexf} from the propagation of random errors (see \citealt{car98-597}). We measured twenty-two atomic and molecular Lick indices in the VIMOS wavelength range (see first column of Table \ref{tab:indices}).

Model comparisons are performed in a relative sense, so we do not calibrate to the Lick instrumental response. However, the line indices do require corrections to account for velocity-broadening and instrumental resolution. Here, we adopt the correction method used in \citet*{smi07-1035}, whereby the correction to zero velocity dispersion and to Lick resolution are performed in one step. Indices measured from a range of stellar population templates, smoothed to the Lick resolution, are compared to those measured from the same templates smoothed to the instrument resolution and further degraded to mimic various levels of velocity broadening. For each index, at a given velocity dispersion, this comparison gives a near linear relation between the measured and corrected index. This technique removes the need to artificially broaden the observed galaxy spectra prior to index measurement, thereby preserving the non-uniform noise properties.

\begin{table}
\centering
\caption{Quantitative comparison of VIMOS data to the previous NFPS line strength measurements. For offset comparisons, CN$_1$, CN$_2$, Mg$_1$ and Mg$_2$ have been converted to \AA{} (from mags).} 
\label{tab:indices} 
\begin{tabular}{@{}crrr} 
\hline
& \multicolumn{1}{c}{offset (\AA{})} & \multicolumn{1}{c}{$\chi^2$ / DoF} & \multicolumn{1}{c}{P($\chi^2$)}\\
& \multicolumn{1}{c}{(1)} & \multicolumn{1}{c}{(2)} & \multicolumn{1}{c}{(3)}\\
\hline
H$\delta$A & --0.33 $\pm$ 0.40 & 9.37 / 9 & 0.60 \\
H$\delta$F & 0.22 $\pm$ 0.22 & 7.98 / 9 & 0.46 \\
CN$_1$ & --0.10 $\pm$ 0.30 & 6.17 / 9 & 0.28 \\
CN$_2$ & --0.04 $\pm$ 0.29 & 4.77 / 9 & 0.15 \\
Ca4227 & 0.14 $\pm$ 0.20 & 19.73 / 9 & 0.98 \\
G4300 & 0.12 $\pm$ 0.18 & 11.31 / 9 & 0.75 \\
H$\gamma$A & 0.32 $\pm$ 0.36 & 15.43 / 9 & 0.92 \\
H$\gamma$F & 0.03 $\pm$ 0.15 & 13.62 / 9 & 0.86 \\
Fe4383 & 0.42 $\pm$ 0.39 & 15.07 / 9 & 0.91 \\
Ca4455 & 0.08 $\pm$ 0.07 & 4.28 / 9 & 0.11 \\
Fe4531 & --0.06 $\pm$ 0.20 & 20.83 / 9 & 0.99 \\
Fe4668 & 0.04 $\pm$ 0.17 & 2.97 / 9 & 0.03 \\
H$\beta$ & 0.13 $\pm$ 0.10 & 15.33 / 9 & 0.92 \\
Fe5015 & 0.48 $\pm$ 0.30 & 27.04 / 9 & 1.00 \\
Mg$_1$ & --0.12 $\pm$ 0.22 & 7.68 / 9 & 0.43 \\
Mg$_2$ & 0.33 $\pm$ 0.17 & 6.50 / 9 & 0.31 \\
Mgb5177 & 0.22 $\pm$ 0.10 & 16.31 / 9 & 0.94 \\
Fe5270 & 0.37 $\pm$ 0.14 & 24.07 / 9 & 1.00 \\
Fe5335 & 0.25 $\pm$ 0.10 & 8.88 / 9 & 0.55 \\
Fe5406 & 0.26 $\pm$ 0.17 & 22.84 / 6 & 1.00 \\
Fe5709 & 0.00 $\pm$ 0.12 & 6.16 / 3 & 0.90 \\
Fe5782 & 0.04 $\pm$ 0.06 & 9.71 / 9 & 0.63 \\
\hline 
\end{tabular}
\begin{flushleft} 
NOTES -- (1) Systematic offset of VIMOS measurements from NFPS. (2) $\chi^2$, given the systematic offset, and degrees of freedom (DoF). (3) Probability that $\chi^2$ is smaller than calculated.
\end{flushleft} 
\end{table} 

\begin{figure}
\includegraphics[viewport=0mm 0mm 75mm 74mm,width=84mm,angle=0,clip]{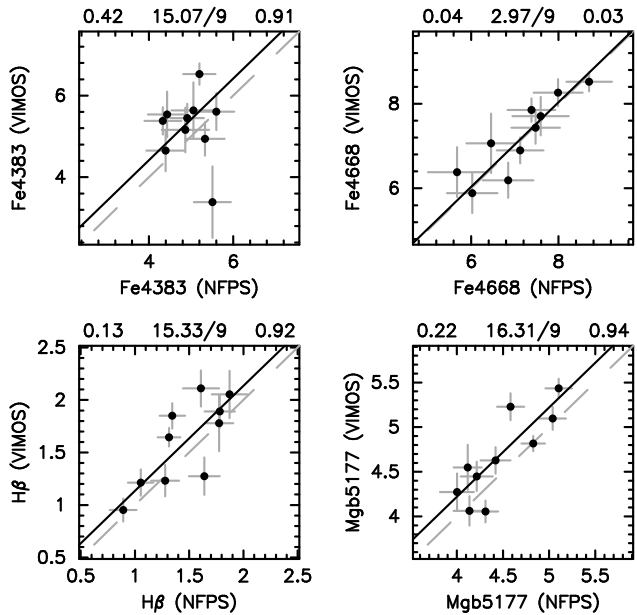}
\caption{Plots showing VIMOS 1 arcsec radius central region absorption line measurements compared with those from NFPS \citep{nel05-137}. Solid line shows median offset; dashed line is equality. Values above each plot show the offset, $\chi^2$ / DoF and P($\chi^2$) respectively, as in Table \ref{tab:indices}. Given the actual scatter, the formal errors on H$\beta$ and Mgb5177 appear under-estimated, whereas for Fe4668 they appear over-estimated.}
\label{fig:nfps}
\end{figure}

Table \ref{tab:indices} compares the central index measurements from VIMOS with those from NFPS for the 10 galaxies in common. The VIMOS measurements are larger in most cases, with an average offset of 0.13 \AA. The $\chi^2$, after accounting for the sample offsets, suggests a reasonable agreement, although the errors on the indices are, in some cases, over- or under- estimated by up to 20 per cent (see Figure \ref{fig:nfps}). 

In Appendix \ref{sec:app_aps} we tabulate aperture corrections derived from the integrated line strengths within two circular apertures (with radii of $r_{\rm e}$ and $r_{\rm e}/3$).

\section{Measurement of gradients}
\label{sec:analysis}

Gradients (for line strengths and stellar population parameters), along with their errors, are calculated using an un-weighted linear least-squares fit to all data points within the effective radius of the galaxy, with radii in log($r/r_{\rm e}$). Our choice of using log-space is not optimal for every galaxy as some targets would certainly be better fit if it were linear, i.e. $r/r_{\rm e}$. The necessity of consistency dictates the use of the same radial units in every case, and the majority of the targets/parameters prefer a fit in log-space. Also in favour of log($r/r_{\rm e}$) is its adoption by previous studies of radial trends. The un-weighted fit ensures sensible results for the minority of cases which would be better suited to linear radial units.

Three of the four smallest galaxies (D48, D53, D61; effective radius $<$ 3 arcsec) have insufficient radial bins with adequate SNR for determining gradients, and as such have not been included in the analysis. The remaining `small' galaxy (D49) has a higher surface brightness, allowing three annuli with SNR $\ge$ 18 within the effective radius.

We first discuss the gradients of the absorption indices (Section \ref{sec:line_grads}). In Section \ref{sec:sp_gen} we calculate the median line strength gradients, from which we infer radial trends in the stellar population parameters, averaged over all galaxies. Finally, a grid inversion technique is used to derive the stellar population gradients for each individual galaxy (Section \ref{sec:ssp_ind}).

\subsection{Line-strength gradients}
\label{sec:line_grads}

The line-strength gradients of the 22 measured indices, for each galaxy in our sample, are tabulated in Appendix \ref{sec:app_lines}.

\begin{figure}
\includegraphics[viewport=0mm 0mm 156mm 177mm,width=84mm,angle=0,clip]{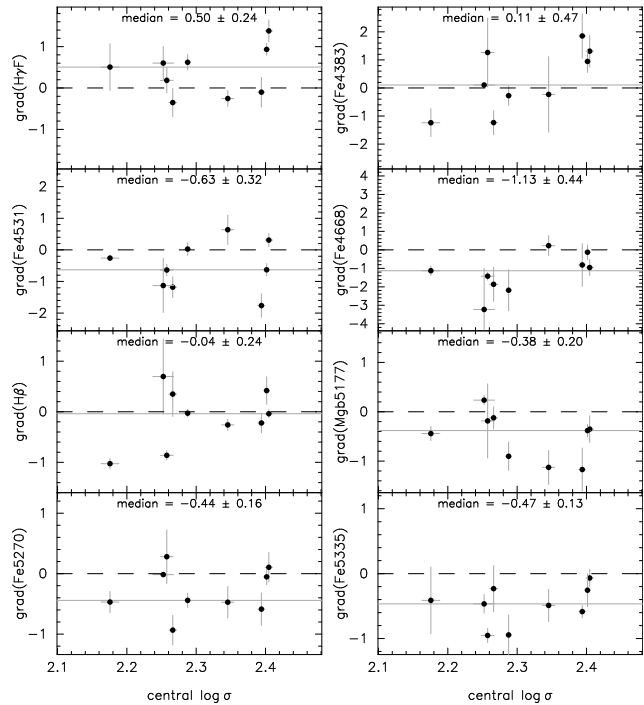}
\caption{Line-strength index gradients versus central log$\,\sigma$. The solid grey line gives the median gradient, which is also displayed at the top of each panel. The dashed black line indicates zero gradient.}
\label{fig:median_grads}
\end{figure}

As an illustration, Figure \ref{fig:median_grads} shows the measured line-strength gradients (against velocity dispersion) for eight of the indices. The median gradient  for the lines is given in each panel. Interestingly, line indices which to first order follow the same physical property, have quite different gradients. For example, the two `age-tracing' Balmer lines show very different gradients: H$\gamma$F generally has a positive gradient (0.50 $\pm$ 0.24) while H$\beta$ has a flat profile (--0.04 $\pm$ 0.24). This can be explained by a higher sensitivity to metallicity of the H$\gamma$F index.

\subsection{Average stellar population gradients}
\label{sec:sp_gen}

By exploring the response of each index to age, metallicity [Z/H] and $\alpha$-abundance [$\alpha$/Fe], we can describe the average radial stellar population trends in our sample.

The linear response of each index to the underlying stellar population parameters were calculated from the \citet{tho03-897,tho04-19} model grids. The responses quantify the change in each index for a decade change in each of age, Z/H and $\alpha$/Fe, derived as described in \citet{smi07-1035}. 

\begin{figure}
\includegraphics[viewport=0mm 0mm 93mm 64mm,width=84mm,angle=0,clip]{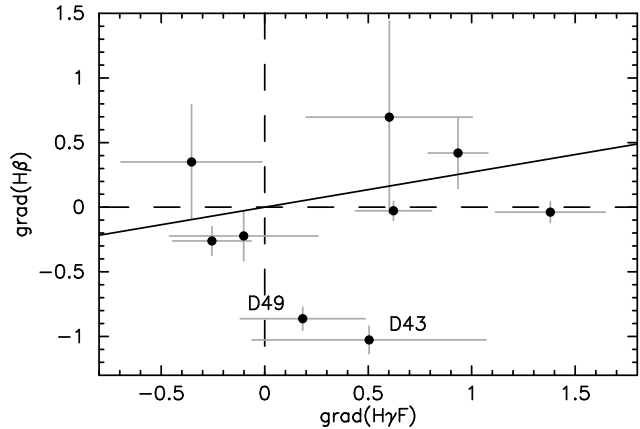}
\caption{H$\beta$ gradient versus H$\gamma$F gradient. The solid line gives the predicted index gradients for a simple model with a [Z/H] gradient only (no radial trend in age or [$\alpha$/Fe]). The two main outliers, presumably the galaxies with strongest age gradients, are labelled.}
\label{fig:balmer_grads}
\end{figure}

Figure \ref{fig:balmer_grads} compares the H$\gamma$F and H$\beta$ Balmer line gradients for each galaxy. Given the derived responses of the Balmer lines, the relation for a metallicity gradient only (no radial trend in age or [$\alpha$/Fe]), is shown. Two galaxies (D43, D49) have a large offset from the predicted Balmer gradients of this simple model. The H$\beta$ response to [$\alpha$/Fe] is too small to explain the discrepancy, requiring a very large negative gradient. This suggests that these galaxies have the strongest age gradients, and given the negative age response of the Balmer lines, the age gradient is expected to be positive. We confirm this interpretation in Section \ref{sec:sp_grads}.

\begin{figure*}
\includegraphics[viewport=0mm 0mm 138mm 66mm,width=120mm,angle=0,clip]{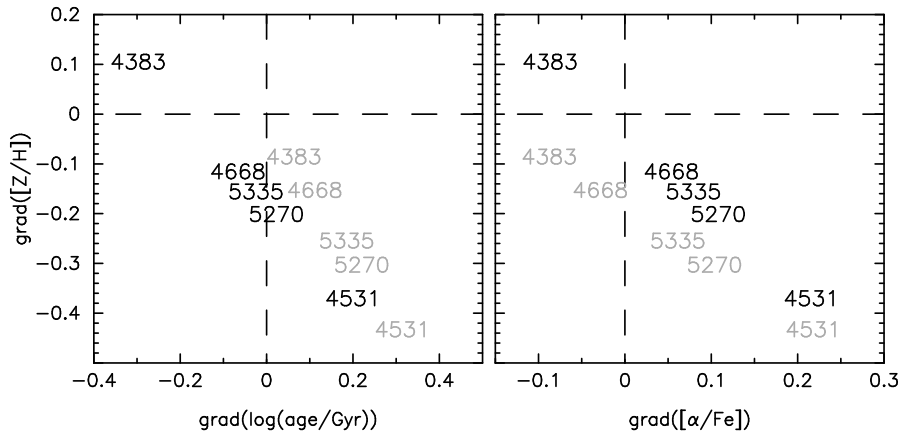}
\caption{Average stellar population gradients {\bf over the nine galaxies}, derived from mean gradients of various triplets of line indices. All triplets incorporate a Balmer line, an iron line and Mgb5177. Points are represented by the iron line wavelength. Black labels indicate H$\gamma$F; grey indicates H$\beta$. The sensitivity of the stellar population gradients to the choice of indices is evident.}
\label{fig:triplets}
\end{figure*}

We consider several triplet combinations of indices, each including a Balmer line, an iron line and Mgb5177. From the mean gradients of the three indices, we use the responses to solve for the average gradients in age, metallicity and $\alpha$-abundance (Figure \ref{fig:triplets}). The left panel indicates that the derived age and metallicity gradients are dependent on the Balmer line/iron line combination, with a scatter in the direction of the usual age--metallicity degeneracy. The right hand panel highlights that the [$\alpha$/Fe] gradients tend to compensate for the age in these models, but also shows that the [$\alpha$/Fe] gradient is generally independent of which Balmer line is used. The triplets including Fe4383 and Fe4531 have the largest uncertainties, due to the large errors for the median line strength gradients used in the stellar population reconstruction.

The dependence of the derived metallicity gradients on the indices used was previously discussed in \citet{san06-823}. These authors found that the index pair Fe4668--H$\beta$ gave a metallicity gradient $\sim$0.4 dex steeper than Fe4383--H$\beta$, a significantly larger difference in metallicity gradient than seen in our equivalent triplets (the index pair plus Mgb5177), where the difference is $\sim$0.07 dex. This underlines that the inclusion of $\alpha$-enhancement via the \citeauthor{tho03-897} models and Mgb5177 yields a higher level of consistency.

The mean stellar population gradients for the sample, as estimated from the triplets of lines, are listed in the first row of Table \ref{tab:mean_sp_grads}. Generally, they support negative metallicity gradients, but flat radial trends for log(age/Gyr) and $\alpha$-abundance.

\begin{table}
\centering
\caption{Mean stellar population gradients. The first two rows show gradients for our VIMOS sample, as derived by two methods: (1) from the median line strength gradients and index responses (Sec. \ref{sec:sp_gen}); (2) individual galaxy grid inversions (Sec. \ref{sec:ssp_ind}). Rows 3 and 4 give gradients from \citet{meh03-423} and \citet{san07-759} respectively, and the final row shows our combined sample (see Sec. \ref{sec:comp}).} 
\label{tab:mean_sp_grads} 
\begin{tabular}{@{}lrrr} 
& \multicolumn{1}{c}{log(age/Gyr)} & \multicolumn{1}{c}{[Z/H]} & \multicolumn{1}{c}{[$\alpha$/Fe]} \\
\hline
This study (1) & 0.07 $\pm$ 0.06 & --0.21 $\pm$ 0.05 & 0.06 $\pm$ 0.03 \\
This study (2) & 0.08 $\pm$ 0.08 & --0.20 $\pm$ 0.05 & --0.02 $\pm$ 0.03 \\
\hline
M03 & 0.03 $\pm$ 0.04 & --0.12 $\pm$ 0.02 & --0.01 $\pm$ 0.01 \\
SB07 & 0.15 $\pm$ 0.10 & --0.31 $\pm$ 0.04 & 0.01 $\pm$ 0.02 \\
Combined & 0.06 $\pm$ 0.03 & --0.18 $\pm$ 0.02 & --0.01 $\pm$ 0.01 \\
\hline 
\end{tabular}
\end{table} 

\subsection{Individual stellar population gradients}
\label{sec:ssp_ind}

We derive stellar population parameters for each radial bin of each galaxy using a grid inversion technique (\citealt*{pro04-1327}; Smith et al. in prep.). This involves a non-linear $\chi^2$ minimisation over several indices, using a cubic-spline interpolation in age--[Z/H]--[$\alpha$/Fe], and a linear extrapolation for points lying outside the model grid.

We use the \citet{tho03-897,tho04-19} models together with eight measured indices: two Balmer lines (H$\gamma$F, H$\beta$), five iron lines (Fe4384, Fe4531, Fe4668, Fe5270, Fe5335) and the $\alpha$/Fe sensitive Mgb5177. The shorter wavelength Balmer indices (H$\delta$A, H$\delta$F) are excluded for having low SNR due to the instrumental response, and H$\gamma$A is redundant as it mostly duplicates the response of H$\gamma$F. The iron lines at the red end (Fe5406, Fe5709, Fe5782) are affected by the 5577 \AA{} sky line and other sky features at the redshift of our cluster.

Stellar population gradients for each galaxy are calculated in an analogous manner to the line strength gradients in Section \ref{sec:line_grads}. The stellar population gradients are listed in Appendix Table \ref{tab:sp_grads_appendix} and plotted for individual galaxies in Figures \ref{fig:age_grads_appendix} -- \ref{fig:afe_grads_appendix}.

\section{Results and discussion}
\label{sec:sp_grads}

The distributions of log(age/Gyr), [Z/H] and [$\alpha$/Fe] gradients for our sample of A3389 galaxies are shown in the top row of Figure \ref{fig:hists}. The mean age and $\alpha$-abundance gradients are consistent with flat radial trends (0.08 $\pm$ 0.08 and --0.02 $\pm$ 0.03 respectively), while the mean metallicity gradient is --0.20 $\pm$ 0.05. The age and metallicity gradients are in agreement with those derived from the mean index gradients (in Sec. \ref{sec:sp_gen}), as summarised in Table \ref{tab:mean_sp_grads}. The $\alpha$-abundance gradients calculated by the two methods marginally disagree. The two galaxies in the most positive age gradient bin (see top row of Fig. \ref{fig:hists}) are D43 and D49, which supports the prediction from the Balmer line plot in Fig. \ref{fig:balmer_grads}.

\subsection{Comparison studies}
\label{sec:comp}

\begin{figure}
\includegraphics[viewport=0mm 0mm 155mm 151mm,height=84mm,angle=270,clip]{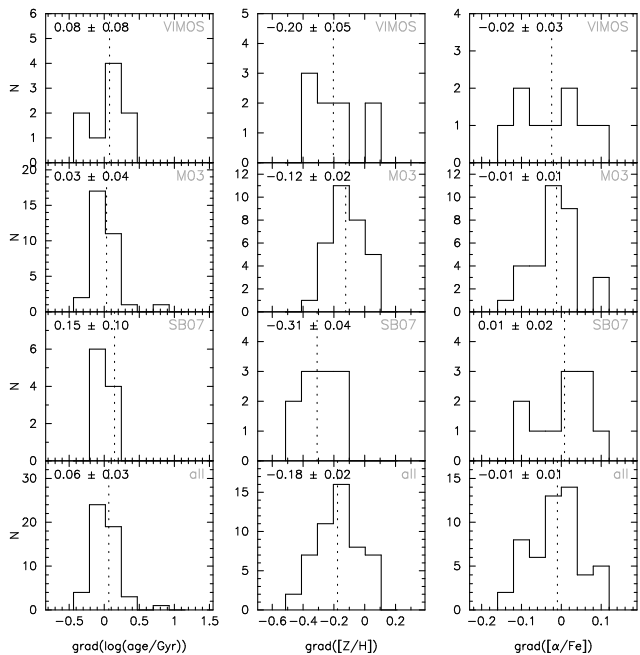}
\caption{Distribution of (left-to-right) log(age/Gyr), [Z/H] and [$\alpha$/Fe]. Shown from top to bottom are our VIMOS data, M03, SB07 and the combined sample (see text for details). Mean gradients are marked with a dotted line and displayed at the top left of each panel.}
\label{fig:hists}
\end{figure}

Stellar population gradients may be affected by numerous variables (environment, morphology etc.), which is likely to result in a large intrinsic scatter within the measured and derived parameters. For small samples, this can contribute to misleading conclusions. We therefore analyse our results alongside data from two previous, long-slit studies of early-type galaxies: \citet[][hereafter M03]{meh03-423} and \citet[][SB07]{san07-759}, chosen due to their use of the same stellar population models. M03 observed a sample of 35 galaxies in the Coma cluster, all morphologically typed as early-type by D80. For their analysis, these authors rebinned the long-slit spectra to achieve a SNR of at least 30 out to the effective radius. Stellar population parameters were derived by an index-pair iteration technique, using H$\beta$, Mgb5177 and $\langle$Fe$\rangle$ = (Fe5270+Fe5335)/2, in conjunction with the \citet{tho03-897} models. SB07 analysed a sample of 11 local early-type galaxies from a mix of environments (field, group, Virgo cluster). The authors rebinned the long-slit spectra along the axis to obtain a minimum SNR $\sim$ 50 out to 2$r_{\rm e}$, and employed the \citeauthor{tho03-897} models, using a $\chi^2$ minimisation technique on 19 Lick indices. The second and third rows of Figure \ref{fig:hists} show the gradient distributions from these studies.

Our results agree well with those from the previous studies (see Table \ref{tab:mean_sp_grads}). In the combined sample, the individual points are not weighted by errors, so the totals are dominated by the larger (but noisy) M03 study. We have not included the early-type galaxy sample from \citet{san06-823} in our comparisons as they use the \citet[][and in prep.]{vaz99-224} stellar population models, which do not explicitly include $\alpha$-enhancement.

We split the combined data from the three samples (VIMOS, M03, SB07) into the two traditional early-type morphology classes (elliptical and S0). Using a Student's t-test to analyse the distributions, there is no significant difference between the classes in any of the stellar population parameters.

\subsection{Stellar population gradients versus central log$\,\sigma$}
\label{sec:grad_sig}

In the following sections, we compare the gradients to central values. The central properties of our VIMOS galaxies are from circular apertures with a radius one third that of the galaxy effective radius. This allows more meaningful galaxy-to-galaxy comparisons than if a fixed radius circle were used in all cases. The long-slit comparison studies measure central values as follows: M03 uses an average along the major axis inside 0.1$r_{\rm e}$/$\sqrt{1-\epsilon}$, where $\epsilon$ is the ellipticity (from optical imaging); SB07 employs an aperture size of 1.5 arcsec $\times$ $r_{\rm e}$/8.

\begin{figure}
\includegraphics[viewport=0mm 0mm 186mm 164mm,height=84mm,angle=270,clip]{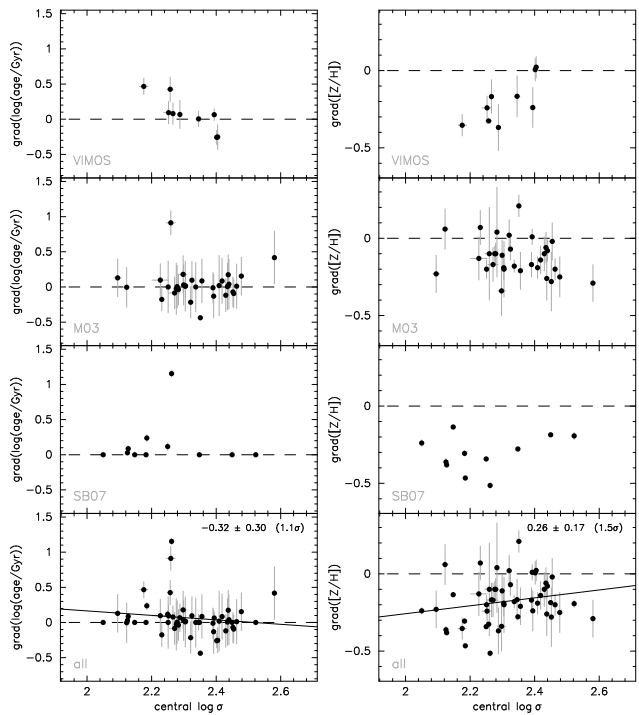}
\caption{Gradient of log(age/Gyr) (left) and [Z/H] (right) versus central log$\,\sigma$. Rows for our VIMOS galaxies, M03, SB07 and the combined sample are shown. Bottom row: solid line is an un-weighted linear best fit to all points; the slope (and significance) is shown in the top right of the panel. The two galaxies with the largest positive age gradients are GMP3201 (from M03) and NGC2865 (SB07).}
\label{fig:sig_grad}
\end{figure}

We explore the relation between stellar population gradients and the central velocity dispersion (log$\,\sigma$, a proxy for galaxy mass). The $\alpha$-enhancement of galaxies suggests the co-requisite of small age gradients \citep[e.g.][]{tho99-655}, while the metallicity gradients are dependent on mergers and feedback processes \citep[e.g.][]{whi80-1,kob04-740}. In a simple model, enrichment gradients depend on the galactic wind efficiency, which varies with the size of the potential well and hence the galaxy mass \citep{ari87-23}. Figure \ref{fig:sig_grad} shows the log(age) (left panel) and [Z/H] (right panel) gradients versus the central log$\,\sigma$. There appears to be no correlation of age gradient with log$\,\sigma$ (slope: --0.32 $\pm$ 0.30). The same is true for the [$\alpha$/Fe] gradient (not shown), which has a slope of --0.04 $\pm$ 0.08 in the combined sample. The [Z/H] gradient weakly correlates with log$\,\sigma$, having a slope of 0.26 $\pm$ 0.17 in the combined sample (although not not statistically significant at 1.5$\sigma$).

SB07 claim a discontinuity at log$\,\sigma$ = 2.3 ($\sigma$ = 200 km s$^{-1}$) in the trend with metallicity, which can be seen in the right-hand panel, third row of Fig. \ref{fig:sig_grad}. They speculate that physically, this could mark the transition between `wet' and `dry' mergers, caused by the interplay between in-flowing gas and the feedback processes (see also \citealt{bin04-1093,fab07-265}). Unfortunately, our VIMOS sample does not include galaxies of low enough mass to test this result. However, the turning point is not seen in the larger, but noisier M03 data (which also dominates the combined plot; bottom row in Fig. \ref{fig:sig_grad}).

\begin{figure}
\includegraphics[viewport=0mm 0mm 53mm 80mm,height=78mm,angle=270,clip]{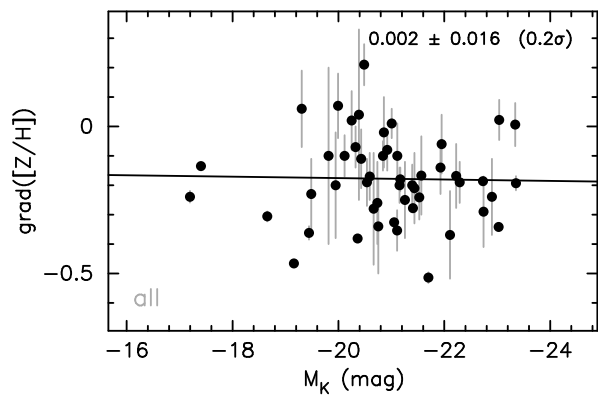}
\caption{Metallicity gradient versus $K$ band absolute magnitude (2MASS) for the combined sample (VIMOS+SB07+M03). Solid line is an un-weighted linear best fit to all points; the slope (and significance) is shown in the top right of the panel.}
\label{fig:kband_all}
\end{figure}

\citet{for05-6}, using a sample from M03 and \citet{san04-phd}, claim a 99 per cent probability of a correlation between [Z/H] gradient and log$\,\sigma$. Additionally, \citeauthor{for05-6} find a 99.6 per cent probability of a correlation between [Z/H] gradient and the $K$-band luminosity (a proxy for stellar mass; from 2MASS), whereas in our combined sample we find no significant trend (Figure \ref{fig:kband_all}).

\begin{figure*}
\includegraphics[viewport=0mm 0mm 186mm 255mm,height=150mm,angle=270,clip]{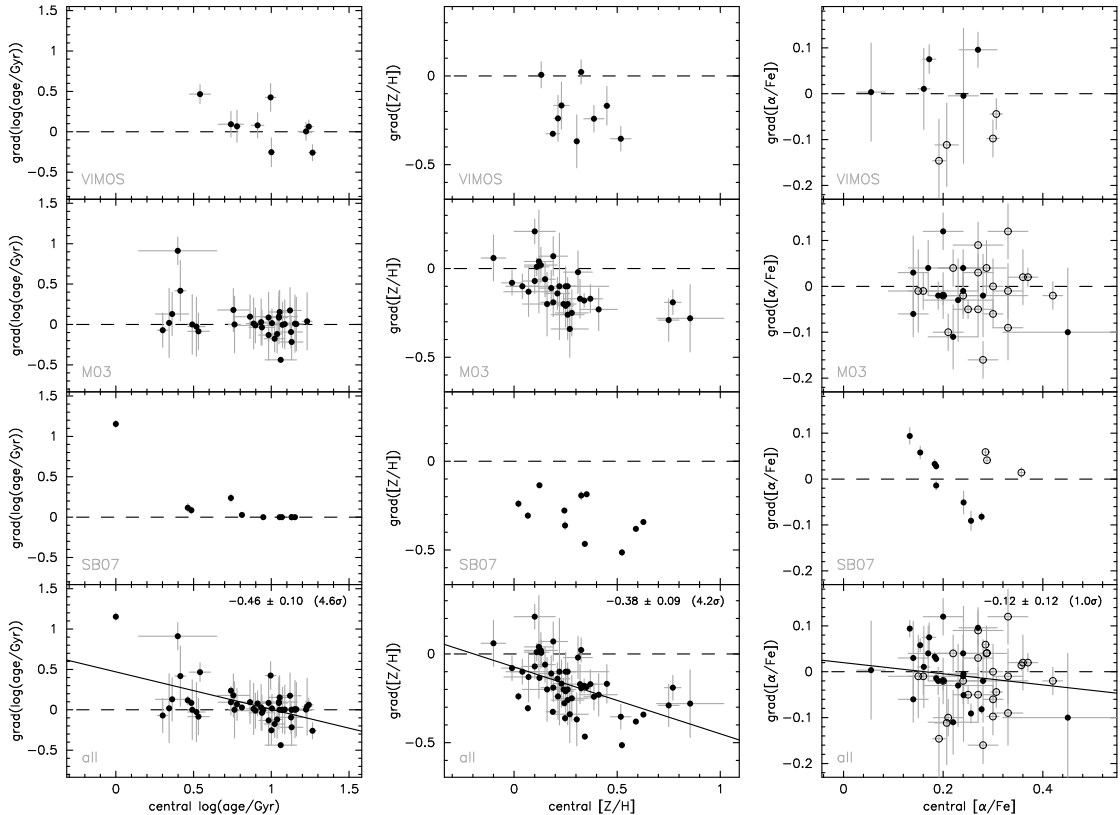}
\caption{Gradient versus central value for log(age/Gyr) (left-hand column), [Z/H] (central column) and [$\alpha$/Fe] (right-hand column). Layout as in Fig. \ref{fig:sig_grad}. Additionally, in the right-hand column, open circles for log$\,\sigma$ $>$ 2.3 else closed circles.}
\label{fig:grad_central}
\end{figure*}

\subsection{Stellar population gradients versus central values}
\label{sec:grad_vals}

In Figure \ref{fig:grad_central} (left-hand panels) we explore the age gradient versus the central value of the age. The bottom panel, which combines the three data sets, shows the best fitting line to the data. The SB07 data illustrates why an un-weighted fit is used throughout this study: several of the galaxies from that study have quoted age gradients of 0.000 $\pm$ 0.000 as a consequence of fitting to discrete model data. In the combined sample, the age gradient has a significant correlation with the central value (4.6$\,\sigma$), with a calculated slope of --0.46 $\pm$ 0.10. The un-weighted fit is dominated by the M03 sample, and the magnitude of the slope may be unduly affected by the two centrally young galaxies with large positive gradients (GMP3201 from M03; NGC2865 from SB07). Without these galaxies, the slope (and significance) is reduced to --0.20 $\pm$ 0.08 (2.4$\,\sigma$).

The central panels of Figure \ref{fig:grad_central} shows the analogous plot for metallicity. The un-weighted linear fit to the combined sample gives a significant trend (4.2$\,\sigma$) with a slope of --0.38 $\pm$ 0.09. This general trend is readily seen in each of the individual datasets, and is noted by SB07 (albeit at 2$\,\sigma$ significance in their small sample). Using central values from an independent study, SB07 conclude that this trend is not due to the correlation of errors. The correlation in our combined sample would be strengthened (slope of --0.56 $\pm$ 0.12; 4.8$\,\sigma$) by removing the three galaxies from M03 (GMP3329, GMP2921, GMP3561) with anomalously metal-rich centres; M03 notes that these galaxies lie outside the model grid leading to large uncertainties in extrapolation. 

Generally, no trends were found between the $\alpha$-abundance gradient and central properties of the galaxy. The right-hand panels of Figure \ref{fig:grad_central} shows the lack of a trend between central [$\alpha$/Fe] and its gradient. SB07 tentatively invoke the dichotomy at log$\,\sigma$ = 2.3 for central $\alpha$-abundance versus the gradient (third row of right-hand panel; open circles for log$\,\sigma$ $>$ 2.3, else closed circles). However, the dichotomy is not recovered in our data, or in M03. The apparent division may be a consequence of the small sample size, with the well known correlation between central velocity dispersion and central [$\alpha$/Fe] \citep[e.g.][]{gra07-243,smi07-1035} ensuring that on average, the high mass galaxies lie further to the right in the plot.

In summary, the observed trends imply that galaxies with young cores have steeper positive age gradients, and those with metal-rich centres have strong negative [Z/H] gradients. Assuming stellar population gradients behave linearly out to the effective radius and continue to do so beyond it, the trends suggest that all galaxies have a similar age and metallicity (old and [Z/H] $\approx$ 0) at $\sim$ 2--2.5 $r_{\rm e}$. Of the 5 galaxies in SB07 which measure [Z/H] to 2 $r_{\rm e}$, four are approximately solar metallicity at this radius and the other is metal-poor. Verification of this result is beyond the scope of our current VIMOS data.

\begin{figure}
\includegraphics[viewport=0mm 0mm 123mm 123mm,height=84mm,angle=270,clip]{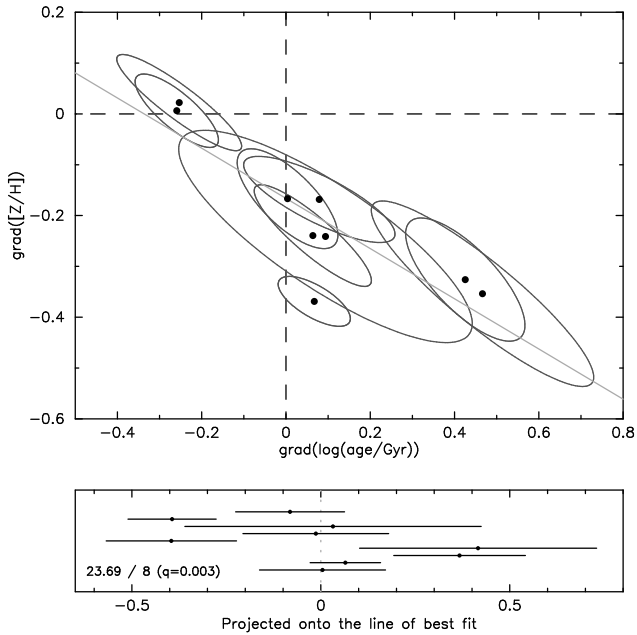}
\caption{Upper panel: measured age and metallicity gradients shown as points, with the 1$\sigma$ confidence ellipses obtained from 10$^3$ Monte Carlo simulations for each of our sample galaxies. Line of best fit shown in grey. Lower panel: one-dimensional distribution of points projected onto the best fit line (shown with an arbitrary y-offset for clarity). Error bars derived from the simulations. The $\chi$-squared / degrees of freedom and Q($\chi^2$) are shown to the lower left. The presence of a real intrinsic scatter is evident.}
\label{fig:error_corr}
\end{figure}

\subsection{Correlation of errors}
\label{sec:corr_errors}

For any particular radial bin, the age and metallicity are not truly independent. The well-known degeneracy between age and [Z/H] is to an extent broken when fitting model spectra to multiple line strength indices. Here, however, the gradients in these parameters suffer from a correlation of errors.

To investigate the importance of correlated errors, we perform 10$^3$ Monte Carlo simulations for each galaxy in our A3389 sample. The realizations consists of a simulated age and metallicity at each radial bin, using a bivariate normal distribution derived from the error ellipse associated with the measured points. The gradients in each parameter are then calculated as with the measured data. Figure \ref{fig:error_corr} (upper panel) shows the measured data as points and the simulated values represented by 1$\,\sigma$ confidence ellipses. The distribution of the points is parallel to the error ellipses, which is reminiscent of the age--mass--metalicity (`Z') plane \citep{tra00-165}.

To test whether any of the scatter is intrinsic, the measured data points are projected onto the line of best fit, using the simulated ellipse major axes as a quantification of the measurement error. Hence, the distribution is reduced to one-dimension along the diagonal (Fig. \ref{fig:error_corr}, lower panel). Taking the null hypothesis of no intrinsic scatter, the $\chi^2$ is 23.69 (8 degrees of freedom) with the probability of obtaining this distribution from the hypothesis calculated as Q($\chi^2$) = 0.003.

Although there is a sizeable correlation of errors, we conclude that there is also a real intrinsic scatter in the same direction. Hence, the correlation we find between the central value and gradient of age is not merely a reflection of the equivalent trend in metallicity, and vice versa. The trends between the central value and gradient in age and metallicity are both real.

\section{Conclusions}
\label{sec:conclusion}

Using the VIMOS integral field unit, we observed a sample of early-type galaxies in the local cluster Abell 3389. From the resulting data-cube, we measured 22 line indices in radial bins out to the effective radius, and successfully derived stellar population gradients. Our key findings are that early-type galaxies have:

\begin{itemize}
\item strong negative gradients in metallicity ($\sim$ --0.2 dex per decade in radius);
\item radial trends in age and $\alpha$-abundance that are consistent with zero;
\item stellar population gradients not significantly correlated with the central velocity dispersion or $K$-band luminosity (proxies for total galaxy and stellar mass respectively);
\item a strong anti-correlation between the central values and gradients of both age and [Z/H]. 
\end{itemize}

In a future paper we will report on an additional sample of 19 early-type galaxies in the Shapley Supercluster, observed with the VIMOS IFU.

\section*{Acknowledgments}
TDR is supported by the STFC Studentship PPA/S/S/2006/04341. RJS is supported by the rolling grant PP/C501568/1 `Extragalactic Astronomy and Cosmology at Durham 2005--2010'. This publication makes use of data products from the Two Micron All Sky Survey, which is a joint project of the University of Massachusetts and the Infrared Processing and Analysis Center/California Institute of Technology, funded by NASA and the National Science Foundation. This research has made use of the NASA/IPAC Extragalactic Database (NED) which is operated by the Jet Propulsion Laboratory, California Institute of Technology, under contract with the National Aeronautics and Space Administration. {\sc iraf} is distributed by the National Optical Astronomy Observatories, which are operated by the Association of Universities for Research in Astronomy, Inc. under cooperative agreement with the National Science Foundation.

\appendix

\section{Derived aperture corrections}
\label{sec:app_aps}

We use the integrated line strengths from two circular apertures, with radii of $r_{\rm e}/3$ and $r_{\rm e}$, to derive aperture corrections. These corrections should aid index line strength translations between apertures of different sizes; for example, between different multi-object spectrographs.

For each index, the median corrections [I($r_{\rm e}$)$\,-\,$I($r_{\rm e}/3$)] over the 9 VIMOS galaxies included in the gradient analysis (i.e. excluding D48, D53, D61) are presented in Table \ref{tab:app_offset}. Corrections are also tabulated for the velocity dispersion (as calculated in Section \ref{sec:sigma}) and the stellar population parameters (as derived in Section \ref{sec:ssp_ind}).

\begin{table}
\centering
\caption{Median index line strength aperture corrections for circular apertures, I($r_{\rm e}$)$\,-\,$I($r_{\rm e}/3$), for the 9 VIMOS galaxies with gradient data (i.e. excluding D48, D53, D61). Values for line indices in \AA{}, except CN$_1$, CN$_2$, Mg$_1$ and Mg$_2$ which are in mags. Q1 \& Q3 give the first and third quartiles of the data. Also tabulated are the equivalent values for the velocity dispersion and stellar population parameters.} 
\label{tab:app_offset} 
\begin{tabular}{@{}crrr} 
\hline
& \multicolumn{1}{c}{median correction} & \multicolumn{1}{c}{Q1} & \multicolumn{1}{c}{Q3} \\
\hline
H$\delta$A & --0.03 $\pm$ 0.48 & --0.91 & 0.91 \\
H$\delta$F & --0.12 $\pm$ 0.34 & --0.38 & 0.53 \\
CN$_1$ & 0.01 $\pm$ 0.01 & --0.03 & 0.03 \\
CN$_2$ & 0.01 $\pm$ 0.01 & --0.02 & 0.02 \\
Ca4227 & 0.01 $\pm$ 0.15 & --0.06 & 0.10 \\
G4300 & --0.03 $\pm$ 0.35 & --0.20 & 0.94 \\
H$\gamma$A & 0.09 $\pm$ 0.33 & --0.53 & 0.37 \\
H$\gamma$F & --0.32 $\pm$ 0.24 & --0.37 & 0.30 \\
Fe4383 & 0.37 $\pm$ 0.35 & --0.29 & 1.01 \\
Ca4455 & --0.07 $\pm$ 0.11 & --0.31 & 0.03 \\
Fe4531 & 0.11 $\pm$ 0.16 & --0.34 & 0.25 \\
Fe4668 & --0.53 $\pm$ 0.33 & --0.99 & --0.29 \\
H$\beta$ & --0.03 $\pm$ 0.09 & --0.19 & 0.08 \\
Fe5015 & 0.04 $\pm$ 0.23 & --0.24 & 0.45 \\
Mg$_1$ & --0.01 $\pm$ 0.01 & --0.01 & --0.01 \\
Mg$_2$ & --0.01 $\pm$ 0.01 & --0.03 & --0.01 \\
Mgb5177 & --0.33 $\pm$ 0.11 & --0.51 & --0.06 \\
Fe5270 & --0.10 $\pm$ 0.19 & --0.21 & 0.05 \\
Fe5335 & --0.12 $\pm$ 0.09 & --0.36 & --0.05 \\
Fe5406 & --0.12 $\pm$ 0.10 & --0.20 & 0.21 \\
Fe5709 & --0.04 $\pm$ 0.29 & --0.14 & 0.21 \\
Fe5782 & --0.07 $\pm$ 0.05 & --0.10 & 0.12 \\
\hline
log$\,\sigma$ & --0.02 $\pm$ 0.02 & --0.04 & 0.00 \\
log(age/Gyr) & 0.03 $\pm$ 0.06 & 0.00 & 0.13 \\
$[$Z/H$]$ & --0.14 $\pm$ 0.04 & --0.17 & --0.09 \\
$[\alpha$/Fe$]$ & --0.04 $\pm$ 0.03 & --0.12 & 0.00 \\ 
\hline
\end{tabular}
\end{table} 

\section{Line strength gradients}
\label{sec:app_lines}

\begin{table*}
\centering
\caption{Absorption index gradients for the galaxies excluding D48, D53, D61. Identifications from \citet{dre80-565}.} 
\label{tab:line_grads_appendix} 
\begin{tabular}{@{}crrrrrr} 
\hline
ID & \multicolumn{1}{c}{H$\delta$A} & \multicolumn{1}{c}{H$\delta$F} & \multicolumn{1}{c}{CN$_1$} & \multicolumn{1}{c}{CN$_2$} & \multicolumn{1}{c}{Ca4227} & \multicolumn{1}{c}{G4300} \\
 & \multicolumn{1}{c}{H$\gamma$A} & \multicolumn{1}{c}{H$\gamma$F} & \multicolumn{1}{c}{Fe4383} & \multicolumn{1}{c}{Ca4455} & \multicolumn{1}{c}{Fe4531} & \multicolumn{1}{c}{Fe4668} \\
 & \multicolumn{1}{c}{H$\beta$} & \multicolumn{1}{c}{Fe5015} & \multicolumn{1}{c}{Mg$_1$} & \multicolumn{1}{c}{Mg$_2$} & \multicolumn{1}{c}{Mgb5177} & \multicolumn{1}{c}{Fe5270} \\
 & \multicolumn{1}{c}{Fe5335} & \multicolumn{1}{c}{Fe5406} & \multicolumn{1}{c}{Fe5709} & \multicolumn{1}{c}{Fe5782} & & \\
\hline
D40 & ~1.016 $\pm$ 0.595 & ~0.532 $\pm$ 0.629 & ~0.611 $\pm$ 0.930 & --0.276 $\pm$ 0.168 & ~0.048 $\pm$ 0.028 & ~0.700 $\pm$ 0.489 \\
 & --0.338 $\pm$ 0.754 & ~0.602 $\pm$ 0.401 & ~0.108 $\pm$ 0.060 & --0.522 $\pm$ 0.258 & --1.127 $\pm$ 0.853 & --3.230 $\pm$ 2.231 \\
 & ~0.697 $\pm$ 0.742 & ~0.448 $\pm$ 0.775 & ~1.034 $\pm$ 0.069 & ~0.955 $\pm$ 0.356 & ~0.235 $\pm$ 0.093 & --0.016 $\pm$ 0.020 \\
 & --0.466 $\pm$ 0.143 & --0.033 $\pm$ 0.729 & --0.161 $\pm$ 0.089 & ~0.151 $\pm$ 0.100 & & \\
\hline
D41 & --2.029 $\pm$ 0.977 & ~0.350 $\pm$ 0.495 & ~1.329 $\pm$ 0.744 & ~1.697 $\pm$ 0.804 & ~0.252 $\pm$ 0.106 & ~2.601 $\pm$ 0.382 \\
 & --1.646 $\pm$ 1.225 & --0.353 $\pm$ 0.341 & --1.235 $\pm$ 0.429 & --0.416 $\pm$ 0.159 & --1.182 $\pm$ 0.334 & --1.861 $\pm$ 0.921 \\
 & ~0.350 $\pm$ 0.441 & ~1.546 $\pm$ 0.670 & --1.154 $\pm$ 0.501 & --0.714 $\pm$ 0.168 & --0.123 $\pm$ 0.225 & --0.934 $\pm$ 0.245 \\
 & --0.232 $\pm$ 0.355 & ~0.183 $\pm$ 0.499 & --0.601 $\pm$ 0.215 & ~0.009 $\pm$ 0.176 & & \\
\hline
D42 & --1.565 $\pm$ 1.607 & --2.049 $\pm$ 1.331 & ~0.740 $\pm$ 0.887 & --0.985 $\pm$ 0.437 & ~1.153 $\pm$ 0.487 & ~0.768 $\pm$ 0.770 \\
 & ~0.618 $\pm$ 0.576 & ~1.379 $\pm$ 0.265 & ~1.312 $\pm$ 0.562 & ~0.184 $\pm$ 0.159 & ~0.308 $\pm$ 0.209 & --0.957 $\pm$ 0.440 \\
 & --0.039 $\pm$ 0.081 & --0.793 $\pm$ 0.218 & --1.008 $\pm$ 0.315 & --0.053 $\pm$ 0.191 & --0.352 $\pm$ 0.272 & ~0.105 $\pm$ 0.245 \\
 & --0.067 $\pm$ 0.135 & --0.597 $\pm$ 0.208 & --0.300 $\pm$ 0.180 & ~0.130 $\pm$ 0.176 & & \\
\hline
D43 & ~1.300 $\pm$ 0.780 & ~0.374 $\pm$ 0.133 & --0.739 $\pm$ 1.018 & --0.620 $\pm$ 1.265 & ~0.390 $\pm$ 0.376 & --1.345 $\pm$ 1.393 \\
 & ~2.456 $\pm$ 1.095 & ~0.504 $\pm$ 0.565 & --1.238 $\pm$ 0.501 & --0.203 $\pm$ 0.115 & --0.260 $\pm$ 0.079 & --1.129 $\pm$ 0.263 \\
 & --1.026 $\pm$ 0.107 & --1.040 $\pm$ 0.397 & --1.876 $\pm$ 1.084 & --0.033 $\pm$ 0.443 & --0.441 $\pm$ 0.146 & --0.470 $\pm$ 0.176 \\
 & --0.412 $\pm$ 0.515 & ~0.222 $\pm$ 0.658 & ~0.001 $\pm$ 0.377 & --0.459 $\pm$ 0.037 & & \\
\hline
D44 & --0.008 $\pm$ 0.500 & --0.510 $\pm$ 0.315 & ~0.162 $\pm$ 0.524 & --0.302 $\pm$ 0.313 & --0.341 $\pm$ 0.072 & ~0.347 $\pm$ 0.349 \\
 & --0.257 $\pm$ 0.348 & --0.255 $\pm$ 0.190 & --0.230 $\pm$ 1.348 & --0.442 $\pm$ 0.277 & ~0.638 $\pm$ 0.469 & ~0.226 $\pm$ 0.541 \\
 & --0.261 $\pm$ 0.111 & --0.337 $\pm$ 1.029 & --1.186 $\pm$ 0.084 & --1.055 $\pm$ 0.804 & --1.128 $\pm$ 0.342 & --0.472 $\pm$ 0.261 \\
 & --0.490 $\pm$ 0.248 & --0.472 $\pm$ 0.266 & ~0.244 $\pm$ 0.068 & --0.550 $\pm$ 0.308 & & \\
\hline
D49 & --2.312 $\pm$ 0.517 & --1.392 $\pm$ 0.008 & ~0.798 $\pm$ 0.613 & ~0.717 $\pm$ 0.805 & --0.133 $\pm$ 0.015 & --0.810 $\pm$ 0.667 \\
 & ~0.365 $\pm$ 0.598 & ~0.184 $\pm$ 0.302 & ~1.263 $\pm$ 1.230 & --0.451 $\pm$ 0.087 & --0.638 $\pm$ 0.182 & --1.424 $\pm$ 0.211 \\
 & --0.862 $\pm$ 0.091 & ~0.019 $\pm$ 0.884 & --0.606 $\pm$ 1.581 & --0.926 $\pm$ 1.189 & --0.186 $\pm$ 0.751 & ~0.281 $\pm$ 0.444 \\
 & --0.952 $\pm$ 0.109 & ~0.443 $\pm$ 0.180 & ~1.178 $\pm$ 0.222 & ~0.232 $\pm$ 0.087 & & \\
\hline
D52 & --0.685 $\pm$ 0.402 & ~0.435 $\pm$ 0.236 & ~0.245 $\pm$ 0.291 & ~0.674 $\pm$ 0.309 & ~0.285 $\pm$ 0.135 & --0.854 $\pm$ 0.553 \\
 & ~1.400 $\pm$ 0.401 & ~0.934 $\pm$ 0.144 & ~0.947 $\pm$ 0.391 & --0.769 $\pm$ 0.235 & --0.630 $\pm$ 0.194 & --0.130 $\pm$ 0.392 \\
 & ~0.419 $\pm$ 0.274 & --0.302 $\pm$ 0.278 & --1.734 $\pm$ 0.289 & --1.190 $\pm$ 0.294 & --0.380 $\pm$ 0.123 & --0.053 $\pm$ 0.130 \\
 & --0.256 $\pm$ 0.253 & --0.693 $\pm$ 0.067 & --1.524 $\pm$ 0.469 & --0.136 $\pm$ 0.052 & & \\
\hline
D60 & ~1.133 $\pm$ 0.606 & ~0.776 $\pm$ 0.393 & --2.291 $\pm$ 0.503 & --2.690 $\pm$ 0.414 & --0.586 $\pm$ 0.439 & --0.373 $\pm$ 0.338 \\
 & ~0.580 $\pm$ 0.302 & ~0.622 $\pm$ 0.185 & --0.275 $\pm$ 0.342 & --0.063 $\pm$ 0.109 & ~0.026 $\pm$ 0.200 & --2.185 $\pm$ 1.117 \\
 & --0.028 $\pm$ 0.074 & --0.494 $\pm$ 0.301 & --1.541 $\pm$ 0.224 & --2.228 $\pm$ 0.782 & --0.901 $\pm$ 0.289 & --0.443 $\pm$ 0.115 \\
 & --0.944 $\pm$ 0.306 & --0.331 $\pm$ 0.106 & --0.264 $\pm$ 0.149 & --0.518 $\pm$ 0.057 & & \\
\hline
D89 & --0.429 $\pm$ 0.672 & --1.220 $\pm$ 0.365 & ~0.686 $\pm$ 0.481 & ~0.679 $\pm$ 0.399 & ~0.561 $\pm$ 0.279 & --0.464 $\pm$ 0.465 \\
 & --1.588 $\pm$ 0.477 & --0.101 $\pm$ 0.359 & ~1.854 $\pm$ 0.792 & ~0.148 $\pm$ 0.151 & --1.762 $\pm$ 0.369 & --0.819 $\pm$ 1.151 \\
 & --0.222 $\pm$ 0.192 & ~1.681 $\pm$ 0.389 & --2.398 $\pm$ 0.545 & --1.716 $\pm$ 0.670 & --1.171 $\pm$ 0.436 & --0.588 $\pm$ 0.268 \\
 & --0.584 $\pm$ 0.096 & --0.040 $\pm$ 0.139 & ~0.206 $\pm$ 0.134 & ~0.158 $\pm$ 0.125 & & \\
\hline 
\end{tabular}
\end{table*}

Table \ref{tab:line_grads_appendix} details the measured kinematics and lines strength gradients for nine of our galaxies (excluding  D48, D53, D61 which have insufficient radial bins for gradient determination; see Section \ref{sec:analysis}).

\section{Stellar population gradients}
\label{sec:app_sps}

\begin{figure*}
\includegraphics[viewport=0mm 0mm 130mm 173mm,height=140mm,angle=270,clip]{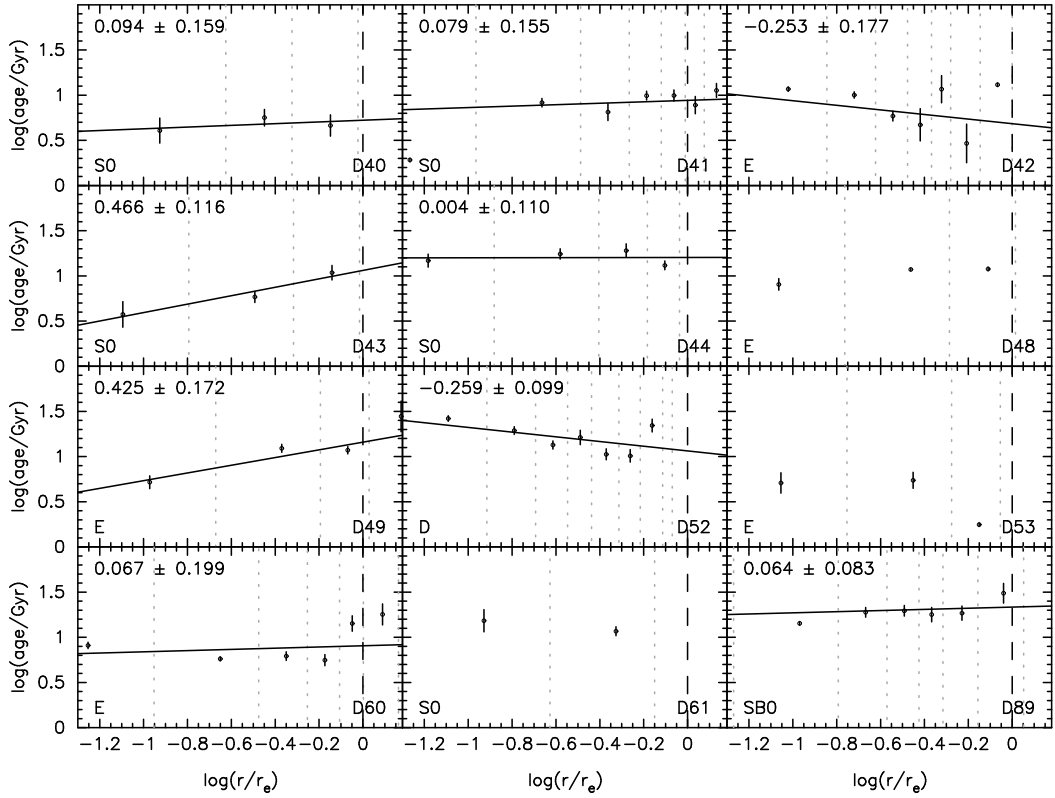}
\caption{log(age/Gyr) versus log($r/r_{\rm e}$). Grey dotted lines show the bin boundaries. Black dashed line is at the effective radius. The fit to points (solid black line) within $r_{\rm e}$ is un-weighted, with the gradient given in the top left of each panel. Name and morphological type \citep[from][]{dre80-565} are given to the bottom right and left respectively. Gradients are not calculated for D48, D53, D61.}
\label{fig:age_grads_appendix}
\end{figure*}

\begin{figure*}
\includegraphics[viewport=0mm 0mm 130mm 173mm,height=140mm,angle=270,clip]{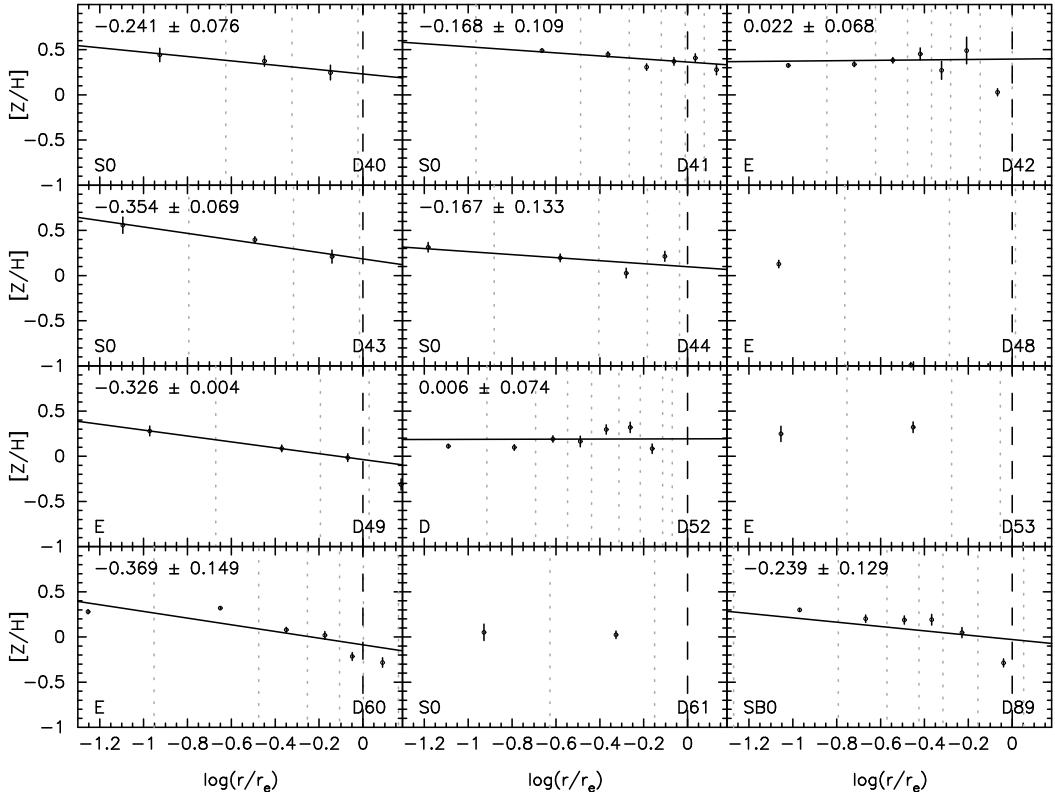}
\caption{[Z/H] versus log($r/r_{\rm e}$). Layout as in Fig. \ref{fig:age_grads_appendix}.}
\label{fig:met_grads_appendix}
\end{figure*}

\begin{figure*}
\includegraphics[viewport=0mm 0mm 130mm 173mm,height=140mm,angle=270,clip]{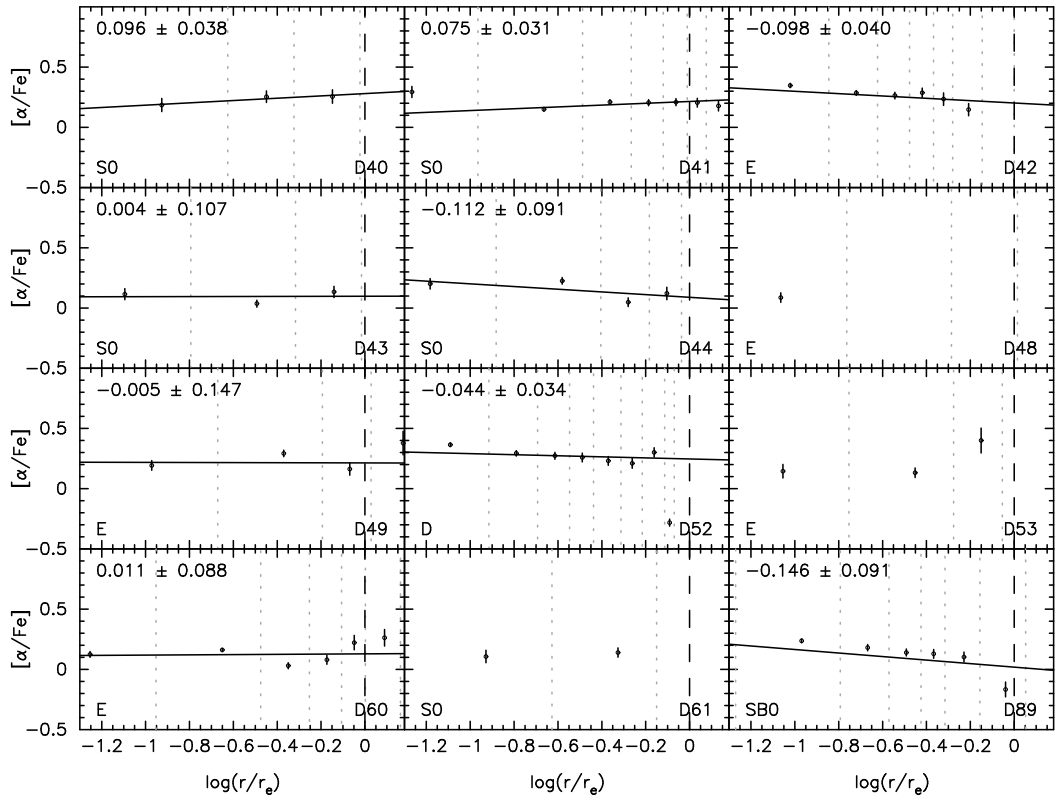}
\caption{[$\alpha$/Fe] versus log($r/r_{\rm e}$). Layout as in Fig. \ref{fig:age_grads_appendix}.}
\label{fig:afe_grads_appendix}
\end{figure*}

Figures \ref{fig:age_grads_appendix}, \ref{fig:met_grads_appendix} and \ref{fig:afe_grads_appendix} show the stellar population parameters versus radius, in log($r/r_{\rm e}$), to the effective radius for all target galaxies. Table \ref{tab:sp_grads_appendix} lists the measured gradients for log$\,\sigma$, log(age/Gyr), [Z/H] and [$\alpha$/Fe], for the nine galaxies analysed.

\begin{table*}
\centering
\caption{Kinematic and stellar population gradients for the galaxies excluding D48, D53, D61. Identifications from \citet{dre80-565}.} 
\label{tab:sp_grads_appendix} 
\begin{tabular}{@{}lcccc} 
\hline
 ID & \multicolumn{1}{c}{grad(log$\,\sigma$)} & \multicolumn{1}{c}{grad(log(age/Gyr))} & \multicolumn{1}{c}{grad([Z/H])} & \multicolumn{1}{c}{grad([$\alpha$/Fe])} \\
 \hline
D40 & ~0.039 $\pm$ 0.018 & ~0.094 $\pm$ 0.159 & --0.241 $\pm$ 0.076 & ~0.096 $\pm$ 0.038 \\
D41 & --0.006 $\pm$ 0.063 & ~0.079 $\pm$ 0.155 & --0.168 $\pm$ 0.109 & ~0.075 $\pm$ 0.031 \\
D42 & --0.051 $\pm$ 0.009 & --0.253 $\pm$ 0.177 & ~0.022 $\pm$ 0.068 & --0.098 $\pm$ 0.040 \\
D43 & --0.059 $\pm$ 0.048 & ~0.466 $\pm$ 0.116 & --0.354 $\pm$ 0.069 & ~0.004 $\pm$ 0.107 \\
D44 & --0.171 $\pm$ 0.010 & ~0.004 $\pm$ 0.110 & --0.167 $\pm$ 0.133 & --0.112 $\pm$ 0.091 \\
D49 & --0.054 $\pm$ 0.038 & ~0.425 $\pm$ 0.172 & --0.326 $\pm$ 0.004 & --0.005 $\pm$ 0.147 \\
D52 & ~0.037 $\pm$ 0.016 & --0.259 $\pm$ 0.099 & ~0.006 $\pm$ 0.074 & --0.044 $\pm$ 0.034 \\
D60 & --0.009 $\pm$ 0.032 & ~0.067 $\pm$ 0.199 & --0.369 $\pm$ 0.149 & ~0.011 $\pm$ 0.088 \\
D89 & --0.164 $\pm$ 0.037 & ~0.064 $\pm$ 0.083 & --0.239 $\pm$ 0.129 & --0.146 $\pm$ 0.091 \\
\hline 
\end{tabular}
\end{table*}

\label{lastpage}

\end{document}